\documentclass[aps,twocolumn,prl]{revtex4}
\usepackage{amsmath}
\usepackage{graphicx}
\usepackage{color}

\begin{document}

\title{Learning the nonlinear interactions from particle trajectories}

\author{Pavel M. Lushnikov$^1$, Petr \v{S}ulc$^2$, and Konstantin S. Turitsyn$^3$}
\affiliation{
$^1$ Department of Mathematics and Statistics, University of New Mexico, Albuquerque, NM, 87131, USA \\
$^2$ Rudolf Peierls Centre for Theoretical Physics, University of Oxford, Oxford OX1 3NP, UK \\
$^3$ Department of Mechanical Engineering, Massachusetts Institute of Technology, Cambridge, MA, 02139, USA
}

\begin{abstract}
Nonlinear interaction of membrane proteins with cytoskeleton and membrane leads to non-Gaussian structure of their displacement probability distribution.
We propose a novel statistical analysis technique for learning the characteristics of the nonlinear potential from the cumulants of the displacement distribution.
The efficiency of the approach is demonstrated on the analysis of kurtosis of the displacement distribution of the particle traveling on a membrane in a cage-type potential.
Results of numerical simulations are supported by analytical predictions. We show that the approach allows robust identification of the potential
for the much lower temporal resolution compare with the mean square displacement analysis.
\end{abstract}

\maketitle


Rapid progress in video-capturing, sub-diffractive microscopes and fluorescence technologies has transformed a single particle tracking (SPT) technology in a powerful tool for studying the properties of biological environments and complex fluids \cite{SaxtonNatMethods2008}. In typical experiment some specific type of biomolecule, i.e. protein or lipid is labeled by  fluorophore or nanoparticle, and its motion is tracked with a camera through in subdiffractive resolution \cite{SaxtonJacobsonAnnuRevBiophysBiomolStruct1997}. The abundance of data available from the SPT experiments has risen demand in data analysis techniques that would help scientists to characterize
the interaction of particle with the environment based on the statistical properties of particle trajectories. Most of the currently used approaches rely on the analysis of second order moments,
like mean square displacement (MSD). The main objective of this Letter is to demonstrate the potential of other statistical properties, that go beyond Gaussian approximation and second-order correlations.
In a practically interesting example of protein moving on the membrane we show that many characteristics of the particle-membrane interactions that can not be recovered from the analysis of MSD
reveal themselves as distinct statistical signatures in the kurtosis of the particle displacement distribution.


The motion of proteins and lipids within biological membranes plays important role in many biological processes. Previous assumption that biological membrane can be considered as two-dimensional fluid
with freely diffusing lipids and proteins \cite{SingerNicolsonScience1972,SaffmanDelbruckPNAS1975}   are now significantly altered by the experimental observations that
membranes are highly heterogeneous \cite{EngelmanNature2005,AlbertsMolecularBiologyCell2007}. Models of lipid rafts, pickets and fences, protein-protein complexes and protein islands were suggested
\cite{LingwoodSimonsScience2010,KusumiAnnRevBio2005,DaumasEtAlBiophys2003,LillemeierEtAlPNAS2006}.
According to these models the motion of the location and diffusion of membrane proteins are significantly influenced by the domains of different
lipid or protein compositions (lipid rafts, protein islands) as well as by the interaction with the cytoskeleton and anchored transmembrane proteins (form fences and pickets, respectively).
E.g., the compartmentalization of the plasma membrane is perhaps the best explained by the fences and pickets \cite{KusumiAnnRevBio2005}. Inside each compartment proteins (lipids)
experience fast diffusion (at time scales $\lesssim 0.01 s$ \cite{KusumiAnnRevBio2005}) which agrees well with the diffusion in the artificial membranes  (which do not have cytoskeleton). A hopping
between different compartments (jump over fences) occurs at  larger time scale $\tau_{h}\sim 0.01 s$. (see e.g. Table 2 in Ref. \cite{KusumiAnnRevBio2005} for the specific values of  $\tau_{h}$
for several types of cells).

Most SPT studies have been relying on the standard video rate ($\sim 30$ frames/sec) \cite{SaxtonNatMethods2008} which does not allow a detailed resolution of the fast diffusion inside compartments because
inter-compartment  hoping rate
$1/\tau_{h}$ exceeds
the video rate. The exception is the work of Kusumi group
\cite{KusumiAnnRevBio2005} which uses 25$\mu s$ temporal resolution.
The analysis of SPT trajectories in the vast majority of previous work has been based on the analysis of MSD \cite{SaxtonNatMethods2008}.
It was demonstrated that SPT with the standard video rate is not sufficient to recover any details about fast diffusion inside compartments as well as any information about compartments \cite{KusumiAnnRevBio2005}.
MSD uses only a small part of information about properties of particle trajectories. The only exception when MSD is optimal corresponds to the pure random walk of the particle when probability distribution
of particle displacement is Gaussian.
However, any inhomogeneity on plasma membrane (represented e.g. by the inhomogeneous potential) results in the non-Gaussianity of that probability distribution which makes MSD non-optimal to recover the properties
of the system from the SPT trajectories.

In parallel to biology SPT based approaches were developed in microfluidics under the name of microrheology \cite{MasonWeitz1995}. Tracking of the Brownian motion of
individual particles immersed in viscoelastic fluid, allows reconstruction of viscoelastic modulus from the Laplace transform of particle MSD.
To our knowledge, all common variations of microrheology (including two-particle and active microrheology) are based on the analysis of second-order correlation functions, and assume linear response of
the viscoelastic fluid. Although microrheological settings are not described in this Letter,
the methods discussed below can be naturally applied there.

In this Letter we propose to  recover major features of the potential from SPT trajectories using the kurtosis as the measure of non-Gaussianity. We demonstrate by the combination of  numerical and analytical methods that
the rate $\sim 100$ frames/sec might be sufficient for
that purpose far  superior to the performance of MSD-based methods. We are not aimed to  fully recover  the spatial distribution of the potential because such type of inverse
problem is ill-posed \cite{FokChouProcRoySoc2010}
and requires very large statistics of SPT trajectories not available in experiment.




The starting point of our work is the following observation. Whenever a particle experiences nonlinear interactions, for example with the cytoskeleton, the probability distribution of the particle displacement
becomes non-Gaussian. Therefore, analysis of the deviations from Gaussian distribution can reveal new information about the nonlinear interactions. There is a infinite number of characteristics that measure the
degree of non-Gaussianity, because the nonlinear interaction can take infinite number of forms. In this Letter we consider only one of the characteristics, that can be accurately estimated with limited
amount of time-series data. Kurtosis of the particle displacement distribution, defined in the following way.
\begin{equation}
 K_\alpha(t) = \frac{\left\langle \Delta r_\alpha(t)^4\right\rangle}{3\left\langle \Delta r_\alpha(t)^2\right\rangle^2} - 1
\end{equation}
Here $\Delta r_\alpha(t) = r_\alpha(t) - r_\alpha(0)$ is the $\alpha$ component of the particle displacement, and the angular brackets denote averaging over the ensemble of particle
trajectories in the same nonlinear environment. Whenever the dynamics of the particle is described by linear equations, the resulting distribution will be Gaussian, and the
kurtosis will be equal to zero for any time interval $t$. It is therefore very different from the extensively studied anomalous diffusion described by MSD $\langle \Delta r_\alpha(t)^2\rangle$ that can be observed even if the dynamics is linear and the distribution is Gaussian. Kurtosis of the displacement distribution originates from the nonlinear interactions of the particle with the environment, and as will be shown below, incorporates a lot of information about the structure of these interactions.

For modeling purposes we consider the motion of Brownian particle (biomolecule) on a membrane with coordinate ${\bf r}\equiv (x,y)$ in a potential $U({\bf r})$ governed by the overdamped  Langevin equation
\begin{equation}\label{langevin1}
\frac{d}{d t}{\bf r}=-\gamma^{-1}\nabla U+\sqrt{2D}\xi(t),
\end{equation}
where $D$ is the diffusion coefficient, $\gamma$ is the friction coefficient and $\xi(t)$ is the random noise with the Gaussian statistics, zero mean and unit variance so
that $\sqrt{2D}\xi(t)$ represents the effect of the thermal noise.




\begin{figure}
\includegraphics[width=0.5\textwidth]{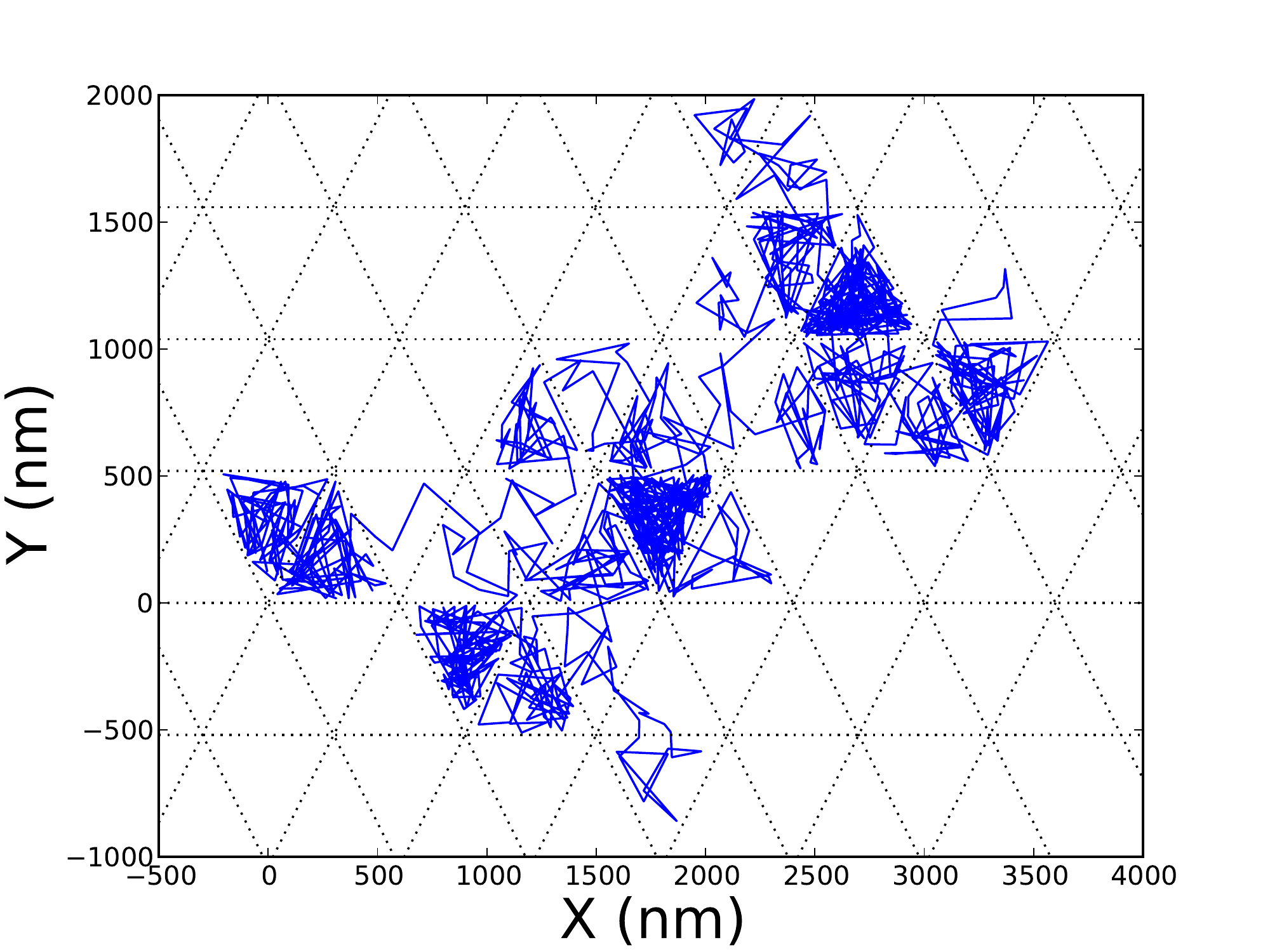}
\centering \caption{Simulation of the trajectory of a diffusing protein in triangular compartments of size $600$ nm during $10$ seconds shown with a time resolution $10$ ms.
The dashed lines indicate the borders between compartments.}
\label{figure_trajectory}
\end{figure}

\begin{figure}
\includegraphics[width=0.5\textwidth]{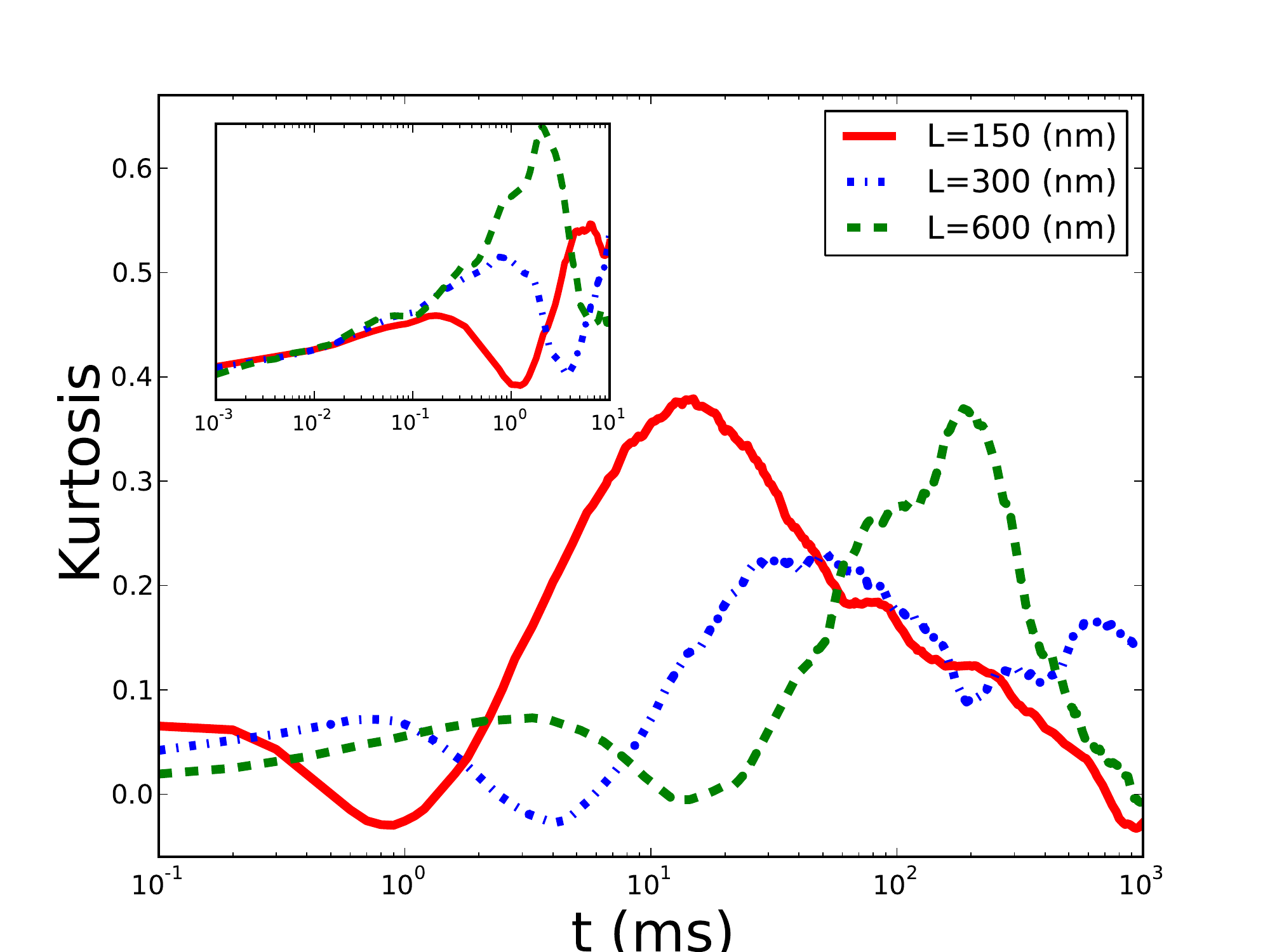}
\centering \caption{The kurtosis of the displacements along $x$ for 3 different compartment sizes calculated with the resolution $0.1 ms$. The inset shows kurtosis scaled as 1/L for  the resolution
$1 \mu s$. }
\label{figure_barriers}
\end{figure}

The equation  (\ref{langevin1}) is simulated using the Monte Carlo algorithm as a two-dimensional (2D) random walk on an elementary triangular lattice composed of
equilateral triangles with side length $a=1$ nm (or $a=0.25$ nm in the case of the inset in Figure \ref{figure_barriers}).

The membrane compartments are typically modeled as equilateral triangles filling entire 2D plane with sides of length $L = 300$ nm, $150$ nm and $600$ nm.
The potential energy on the elementary lattice is labeled as  $U_i=U({\bf r}_i)$, where $i$ is 2D index on that lattice.
Figure \ref{figure_trajectory} shows the typical simulation of total time $10$ s.
In each simulation step, a random closest neighbor $j$ of site $i$ that is occupied by the random walker (diffusing protein) is chosen.
The move to site $j$ from $i$ is accepted with probability
 $p = \min (1,\exp{[ ( U_i - U_j )/T]}). $
We set the temperature $T = 1$  which implies by the Einstein relation $D=T/ \gamma$ that $\gamma=1/D$ in  (\ref{langevin1}).
The potential of barriers between compartments is defined as
$ U_i(l) =  H \exp{ ( -  d_{i,l}^2/\sigma^2   )},
$
where $U_i(l)$ is the the contribution to the total potential $U_i$ at site $i$ from $l$th barrier and $d_{i,l}$ is the distance from site $i$ to $l$th barrier. $U_i$ is given by the sum of contributions from all barriers:
 $U_i=\sum_l U_i(l)$. Also $H$ is the height of the barrier and the
width of the barrier is $2 \sigma$. By default  the barrier parameter $\sigma$ is set to $5$ nm and the height is set to $H=7$.

We set the diffusion coefficient to $D = 2.5 \mu m^2 s^{-1}$ on the lattice, which means that one iteration step of the MC simulation corresponds to a
time period $\tau = 0.1 \mu s$ for lattice size $a=1 nm$ and $\tau = 0.00625 \mu s$ for  $a=0.25 nm$ according to the relation $D=a^2/(4\tau)$.
Effect of discretization is negligible at time scales $\gg \tau$ which motivates our choice of the numerical values of $a.$


In absence of the potential $U$ the particle experiences a Brownian motion with $\langle\Delta x^2+\Delta y^2\rangle = 4 D t$ and $K(t) = 0$.
However, if $U$ is nonzero, the kurtosis of the particle motion acquires a non-trivial shape as shown in Figure \ref{figure_barriers} for  three different compartment sizes.
The typical length of  simulation was $10^9$ steps with every $10^3$-th step  recorded, thus corresponding to an experiment of total length $100
s$ with a time resolution (i.e. inverse frame rate) $0.1$ ms.
Note that we always use the elementary time step $\tau$ to generate particle trajectories. Experimental observations have much lower temporal resolution than $\tau$ and to imitate such resolution
we use only a small fraction of simulation points to calculate the kurtosis for each chosen resolution.
  Figure \ref{figure_barriers} shows that the kurtosis is characterized by two peaks separated by a local minima.
Note that the position of the minima scales as $~L^2$.
The inset of Figure \ref{figure_barriers} shows a simulation with lattice size $a=0.25 nm$,  the duration $10 s$ and  the resolution $1 \mu s$, which we used to test the analytical
predictions for kurtosis behavior
for low times as discussed below. The kurtosis in that inset is divided by $L$. The kurtosis scales as $1/L$ for low times $t$ and grows as $t^{1/2}$, as expected from our analysis.

\begin{figure}
\includegraphics[width=0.5\textwidth]{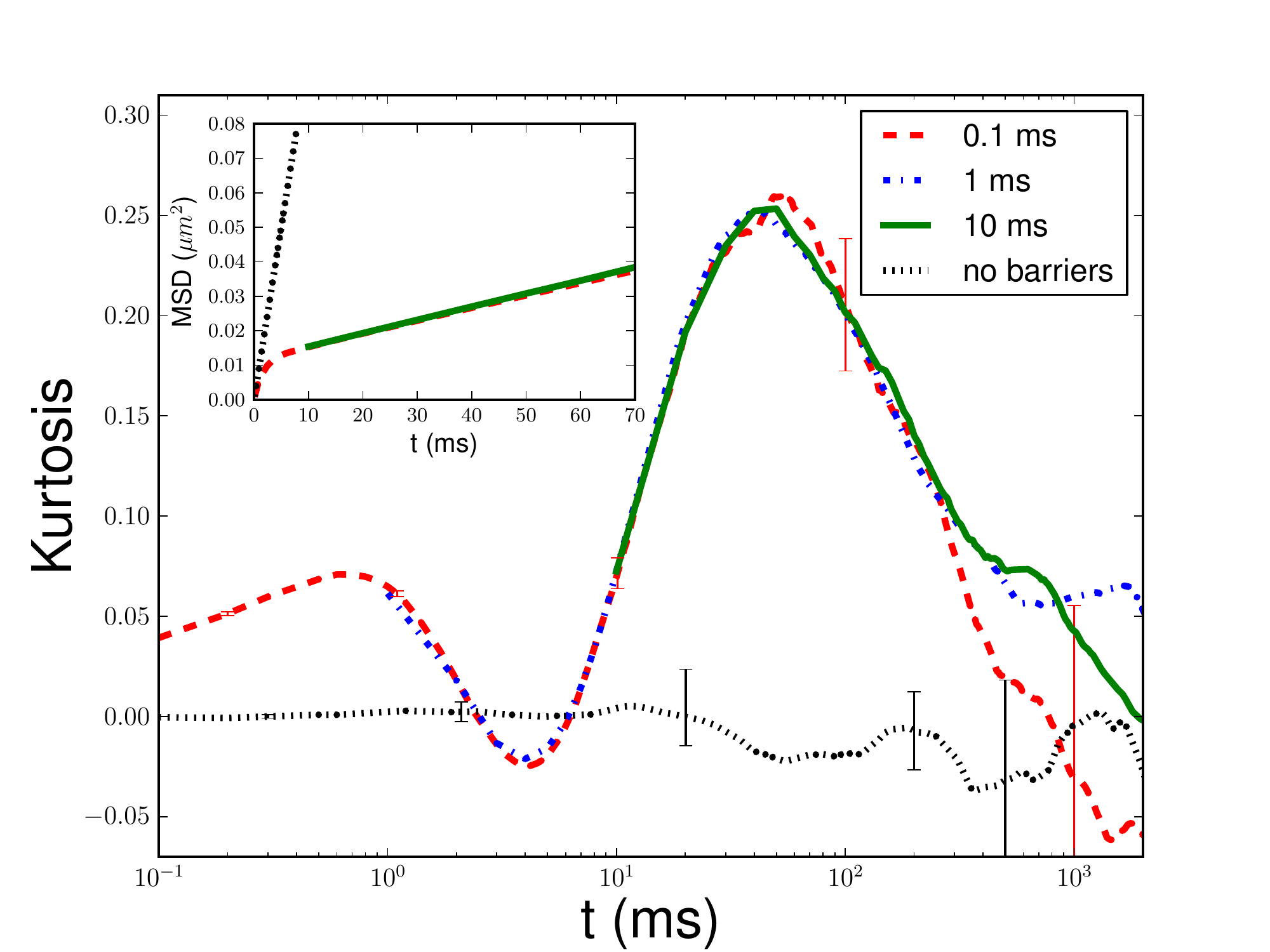}
\centering \caption{The kurtosis along $x$ for 3 different resolutions with $L=300$ nm.
The kurtosis for pure diffusion (no barriers) is also shown with the resolution $0.1 ms.$ Error bars correspond to   $0.1 ms$ resolution.
The inset shows MSD for diffusion with no compartments (short dashed line) and for simulations with the resolutions $0.1 ms$ (long dashed line) and $10 ms$ (solid line).}
\label{figure_videorates}
\end{figure}

The kurtosis of simulations with three different temporal resolutions  $0.1 ms$, $1 ms$ and $10 ms$ for the compartment size $L=300 nm$ are shown in Figure \ref{figure_videorates}.
Already for the resolution $\simeq 10 ms$ it is possible to see the characteristic features of the kurtosis inferring compartments' structure by comparing with the pure diffusion case.
The kurtosis curves are averaged over 5 different simulation runs,
each of duration $100 s$ for the resolution $0.1 ms$ and $500 s$ for the resolutions $1 ms$ and $5 ms$. We also show the kurtosis for pure diffusion simulations, which should be equal to 0
for the infinite trajectory.
The error bars show standard deviation for the
five simulations with  the resolution $0.1 ms$.
The inset compares MSD of pure diffusion with no barriers and two different simulations with the resolutions $0.1 ms$ and $10 ms$ for $L=300$nm. While the MSD plot with the resolution $0.1 ms$
shows transition from the fast diffusion regime inside the compartment with  $D = 2.5 \mu m^2 s^{-1}$ to the slow hopping diffusion regime with $D = 0.1 \mu m^2 s^{-1}$, the MSD plot of $10 ms$ resolution
is able to capture only the slow diffusion.

\begin{figure}
\includegraphics[width=0.5\textwidth]{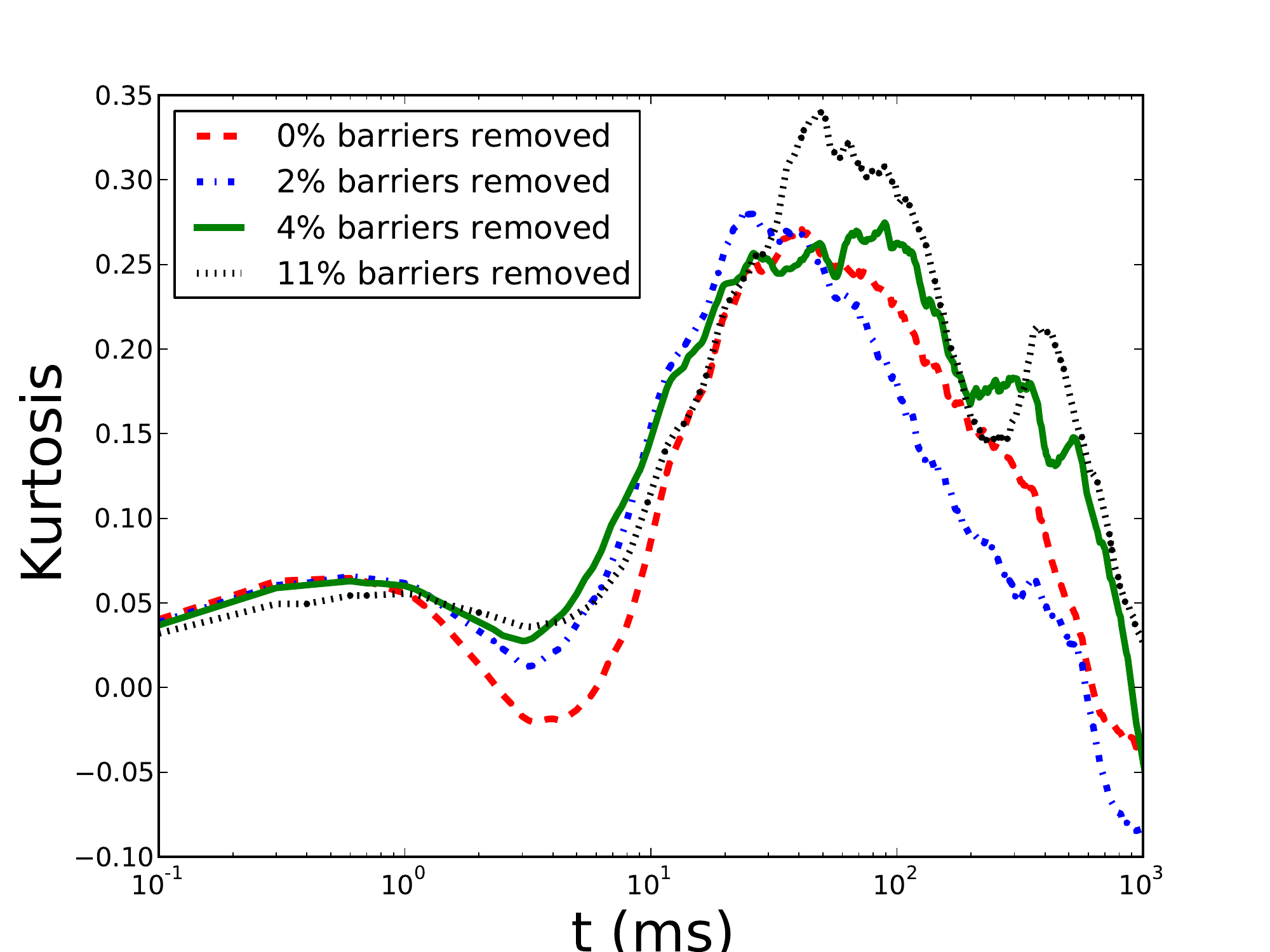}
\centering \caption{The kurtosis along $x$ for  different fractions of randomly removed barriers for simulation of 100s duration. 
}
\label{figure_barriersremoved}
\end{figure}

Figure \ref{figure_barriersremoved} shows that the kurtosis is quite robust to the defects of the triangular compartment lattice. We also found similar robustness when we randomly distorted equilateral triangular lattice by
10\% in angles compare with $\pi/3$ angles as well as when we used rectangular lattice instead of triangular one.






Observed structure of the kurtosis can be explained theoretically in a limiting case when the characteristic time $\tau_d \sim L^2/D$ associated with the diffusion inside the compartment is much smaller than the hopping time that estimates the interval between consecutive hopping over the barrier: $\tau_h \sim \tau_d\exp(\Delta U/T)$, where $\Delta U$ is the height of the barrier.


The initial rise of the kurtosis function $K(t)$ in the region $t\ll
\tau_d$ is related to short trajectories that were reflected from the
boundary. In order to understand this rise qualitatively one can analyze
a simplified problem of one-dimensional diffusion in the neighborhood
of reflecting wall. Assuming that the random walk takes place in the
$x$ direction, the Green function which is defined as a probability
density of the final position $x=X(t)$ assuming that the particle was
at position $x_0$ at $t=0$ is given by $G(x;x_0,t) =
G_0(x-x_0;t)+G(x+x_0;t)$, where $G_0(x;t) = (2\pi D t)^{-1/2}
\exp\left[-x^2/2Dt\right]$. In the above we assumed that the
reflecting boundary is located at $x=0$. The kurtosis can be found by
calculating the expressions for moments $C_n = \int_0^\infty dx
(x-x_0)^n G(x;x_0,t)$ with $n=2,4$. The moments have to be then
averaged over the value of $x_0$, that we assume to be uniformly
distributed in the region $0<x<L$ with $L\gg \sqrt{Dt}$ being a
characteristic compartment size. Straightforward integration yields
the following expressions for the moments in the leading order over
$\sqrt{Dt}/L$: $C_2 = Dt/2 - (Dt)^{3/2}/3\sqrt{\pi} L$ and $C_4 = 3
(Dt)^2/4 - 4 (Dt)^{5/2}/5 \sqrt{\pi}L^2$. Note that the correction to
$C_4$ is smaller in comparison to the correction to $C_2^2$ which
explains the initial rise of kurtosis $K = 4 (Dt)^{1/2}/15\sqrt{\pi}
L$. The numerical factors in this expression are not
universal and may depend on the actual form of the
compartment. However, the scaling laws $K \propto t^{1/2}$ and
$K\propto L^{-1}$ are universal, they are seen in the inset of  Figure \ref{figure_barriers} and can be checked
experimentally.

For intermediate timescales $\tau_d \ll t \ll \tau_h$ the particle has
enough time to diffuse around its compartment, however the events of
passing through the barrier are still rare. In this regime one can
calculate the value of kurtosis by assuming that the Green function
$G(\bm{r};\bm{r}_0,t) = P_\infty(\bm{r})$, where $P_\infty(\bm{r})$ is
the uniform distribution with support inside the compartment. This
assumption implies that the particle had enough time to explore the
whole compartment, and moreover, that the width of the barriers is
negligible in comparison to the compartment size. If the later
approximation is not justified, the Green function has to be replaced
by equilibrium Boltzmann-type distribution $G(\bm{r};\bm{r}_0,t) =
\exp[-U(\bm{r})/T]/Z$.
As long as the initial particle position is also
uniformly distributed over the compartment we obtain the following
expressions for the moments of particle jumps: $C_n = \int_C d{\bm
r}\int_C{\bm r}_0 ({\bm r}-{\bm r}_0)^n
P_\infty(\bm{r})P_\infty(\bm{r}_0)$ that yields $K(t)=-1/10$ for
triangular compartments with very thin barriers. Note, that the value
of kurtosis on these timescales is sensitive to the actual shape of
compartments and to the width of the barrier potentials and can
therefore be used for getting experimental insight on these membrane
properties.

The late time asymptote $t\gtrsim \tau_h$, that is responsible for the second peak
of the observed kurtosis is determined by trajectories that had a finite number of hops between the
compartments. The non-gaussianity of the jump distance distribution is
related to the discrete nature of barrier hopping events. When the typical number of hopping events is small, say $2-3$,
the fluctuations of the total distance traveled by a particle are stronger than Gaussian, and that explains the rise of the kurtosis at $t\approx \tau_h$. As the number of hops becomes very large for $t \gg \tau_h$, the distribution becomes Gaussian again, due to central limit theorem. The kurtosis decays back to zero. Analytical results for the kurtosis behavior in this region will be reported
elsewhere.



To conclude, we have proposed and analyzed a novel particle-tracking approach for identification of nonlinear interactions. Unlike common MSD techniques, our approach is based on the analysis of non-Gaussian characteristics of the particle dynamics, specifically the kurtosis of the displacement probability distribution.
The functional dependence of the kurtosis on the measurement time carries a lot of information about the nonlinear interactions that contribute to the particle motion.
E.g., if a particle is placed in a cage-type potential induced by
cytoskeleton or transmembrane proteins, the resulting kurtosis of the displacement is a non-monotonous function with three distinct regions characterized by the change of the sign of the kurtosis slope.
 Specific structure of the kurtosis function depends on the characteristics of the potential: shape and size of the individual cells, heights and widths of the barriers but the change of sign feature is quite robust to the specifics
 of the potential.

A number of other techniques have been proposed for identification of the potentials on biological membranes based on the analysis of individual trajectories.
 Most comprehensive approach is to solve the inverse problem of reconstruction of $U({\bf r})$ from  the  trajectories. However, such problem is generally ill-posed \cite{FokChouProcRoySoc2010}
 and needs very large statistics of trajectories. E.g.,
Ref.\cite{MassonVergassolaEtAlPRL2009} suggested to infer  forces acting on the biomolecule
\cite{AuthGovBiophysJ2009}
requiring the multiple particle visits of each spatial location which is difficult to achieve experimentally.
In addition, the potential $U$ in living cells can slowly change with time. This fact may limit the application of inverse problem approaches,
that attempt to reconstruct the specific potential.
 Another approach \cite{KusumiSakoYamamotoBiophysJ1993,KusumiAnnRevBio2005} focuses on identification of the potential barriers from MSD-based analysis. That approach is successful but requires very high temporal resolution of trajectories.
 One more method is based on the measurement of autocorrelation
of SPT trajectories and recovering of the probability distribution of particle jumps \cite{YingHuertaSteinbergZunigaBullMathBiol2009} which indirectly displays the information about the inhomogeneity of
the plasma membrane.
In contrary, the technique analyzed in this Letter has more modest goals of learning the characteristic scales associated with the potential.
It does not rely on long observations of individual particles, and can be based on aggregation of time-series from an ensemble of measurements of different particles in the same class of membranes.











Work of P.L. was supported by NSF grant DMS 0719895.


\begin{thebibliography}{98}

\bibitem{SaxtonNatMethods2008} M. J. Saxton, Nat. Methods {\bf 5}, 671 (2008).

\bibitem{SingerNicolsonScience1972} S.J. Singer, and G.L. Nicolson.  Science {\bf 175}, 720 (1972).

\bibitem{SaffmanDelbruckPNAS1975} P.G. Saffman, M. Delbr{\"u}ck. Proc.
Natl. Acad. Sci. USA {\bf 72}, 3111 (1975).

\bibitem{SaxtonJacobsonAnnuRevBiophysBiomolStruct1997} M. J. Saxton, and K. Jacobson. 
Annu. Rev. Biophys. Biomol. Struct., {\bf 26}, 373 (1997).

\bibitem{EngelmanNature2005} D.M. Engelman, Nature, 
{\bf 438}, 578 (2005).

\bibitem{AlbertsMolecularBiologyCell2007} B. Alberts, {\it Molecular Biology of the Cell, 5 edition} (Garland Science, NY, 2007).


\bibitem{LingwoodSimonsScience2010} D. Lingwood, and K. Simons, Science, {\bf 327}, 46 (2010).

\bibitem{KusumiAnnRevBio2005} A. Kusumi, {\it et. al.}, Ann. Rev. BioPhys. Biomol Struct., {\bf 34}, 351 (2005).

\bibitem {DaumasEtAlBiophys2003} F. Daumas, {\it et. al.,} Biophys. J., {\bf  84}, 356 (2003).

\bibitem{LillemeierEtAlPNAS2006} B. F. Lillemeier, J.R. Pfeiffer, Z. Surviladze, B.S. Wilson, and M.M. Davis, Proc. Nat. Acad. Sci., {\bf 103}, 18992 (2006).

\bibitem{FokChouProcRoySoc2010} P.W. Fok, and T. Chou, Proc. Roy. Soc. A, {466},  3479 (2010).

\bibitem{MasonWeitz1995} T.G. Mason, and D.A. Weitz, Phys. Rev. Lett. {\bf 74}, 1250 (1995).

\bibitem{MassonVergassolaEtAlPRL2009}  J.-B. Masson, {\it et. al.}  
Phys. Rev. Lett., {\bf 102}, 048103 (2009).

\bibitem{AuthGovBiophysJ2009}  T. Auth, and N.S. Gov,  
Biophys. J., {\bf 96}, 818 (2009).

\bibitem{KusumiSakoYamamotoBiophysJ1993}
A. Kusumi, Y. Sako, and M. Yamamoto,
Biophys. J. {\bf 65}, 2021 (1993).


\bibitem{YingHuertaSteinbergZunigaBullMathBiol2009}
W. Ying, G. Huerta, S. Steinberg, and M. Z{\' u}{\~n}iga, Bull. Math. Biol., {\bf 71}, 1967 (2009).


\end{thebibliography}
\end{document}